\documentstyle[twocolumn,prl,aps]{revtex}
\large
\begin{document}
\draft
\twocolumn[
\hsize\textwidth\columnwidth\hsize\csname @twocolumnfalse\endcsname

\title
      {
        Non-dynamic origin of the acoustic attenuation at high frequency
        in glasses
      }
\author{        
        G.~Ruocco$^{1}$,
        F.~Sette$^{2}$,
        R.~Di Leonardo$^{1}$,
        D.~Fioretto$^{3}$,
        M.~Lorenzen$^{2}$,\\
        M.~Krisch$^{2}$,
        C.~Masciovecchio$^{2}$,
        G.~Monaco$^{1}$,
        F.~Pignon$^{2}$,
        T.~Scopigno$^{4}$
                }
\address{
         $^1$
         Dipartimento di Fisica and INFM, Universit\`a di L'Aquila,
	 I-67100, l'Aquila, Italy. \\
         $^2$
         European Synchrotron Radiation Facility, BP 220,
         F-38043, Grenoble Cedex, France. \\
         $^3$
         Dipartimento di Fisica and INFM, Universit\'a di Perugia,
	 I-06100, Perugia, Italy.\\
         $^4$
         Dipartimento di Fisica and INFM, Universit\'a di Trento,
	 I-38050, Povo, Trento, Italy.\\
        }

\date{\today}
\maketitle
\begin{abstract}
The sound attenuation in the {\it THz} region is
studied down to $T$=16 K in glassy glycerol
by inelastic x-ray scattering. At striking variance with the 
decrease found below $\approx$100 K in the {\it GHz} data, the 
attenuation in the {\it THz} range does not show any $T$ 
dependence. This result {\it i)} indicates the presence of 
two different attenuation mechanisms, active respectively in 
the high and low frequency limits; {\it ii)} demonstrates the 
non-dynamic origin of the attenuation of {\it THz} sound waves, 
and confirms a similar conclusion obtained in $SiO_2$ glass 
by molecular dynamics; and {\it iii)} supports the low
frequency attenuation mechanism proposed by 
Fabian and Allen (Phys.Rev.Lett. {\bf 82}, 1478 (1999) ).
\end{abstract}


]

One of the most important and still unsettled subjects in the physics of
topologically disordered systems regards the mechanisms for the propagation
and attenuation of density fluctuations. The propagating nature of
acoustic waves, as seen by Ultrasonic and Brillouin Light Scattering (BLS) 
measurements in the $MHz$ and $GHz$ region respectively, has been shown to 
persist up to the {\it THz} region by the existence of a linear relation 
between the peak energy, $E$, and the momentum transfer, $Q$, of the
inelastic features observed in the dynamic structure factor, $S(Q,E)$,
of glasses \cite{sci}. This result is the outcome of extensive studies on 
the shape of $S(Q,E)$ performed using Molecular Dynamics (MD) simulations 
\cite{siomd,MD1,MD2,MDsioE} and the newly developed Inelastic X-rays 
Scattering (IXS) technique \cite{sci,IXS1,IXS2,IXS3}. This latter technique
allows to study the $S(Q,E)$ in the "high" $Q$ range ($Q \approx$1-10
nm$^{-1}$), thus increasing by about two orders of magnitude the $Q$
values typically investigated by BLS ($Q \approx$0.01-0.04 nm$^{-1}$).
In the IXS and MD $Q$-range, beside the persistence of a linear dispersion
of the acoustic excitation energies, one also observes a progressive
broadening of the inelastic features, which is responsible for their
disappearance at a certain $Q_m$ value. Typically $Q_m$ is some tenths
of $Q_M$ $-$ the position of the first sharp diffraction peak in the static
structure factor, $S(Q)$ \cite{sci}. The study of the mechanisms leading to
this damping, and, therefore, the investigation of the sound waves
attenuation at these $Q$-values $-$ characteristic of structural
correlations at the interparticle level $-$ is obviously of great interest. 

The acoustic excitations at frequencies in the $THz$ range, as measured so
far in glasses and glass forming liquids by IXS, have a linewidth parameter
$\Gamma_Q$ which seems to show a $Q^2$ dependence \cite{sci}. Moreover $-$
in all the IXS data reported so far $-$ $\Gamma_Q/Q^2$ has a negligible
temperature dependence in a wide temperature region ranging from values
well below the glass transition temperature, $T_g$, up to the liquid phase
\cite{sci}. At variance with this behavior, as well known, the linewidth
of the excitations in the $GHz$ region, measured by BLS, show a relevant
temperature dependence, which becomes particularly strong in the limit
of very small temperatures
\cite{BLSgly,BLSsio1,BLSsio2,BLSsio3,BLSsio4,BLSbo}. The temperature
dependence of the linewidth in the $GHz$ range has motivated many
theoretical studies, leading to different hypotheses on the frequency
(or $Q$) evolution of the attenuation mechanisms \cite{BLShun}.

In this Letter we report an IXS study on the low temperature behavior of
the excitations linewidth in glassy glycerol. Specifically, we concentrate
on the study of $THz$ excitations in the temperature region where the BLS
data in the $GHz$ range show a marked temperature variation. Within the
error bar, the linewidth measured by IXS is {\it temperature-independent}
in the whole $0.1 T_g$ to $T_g$ region, whereas, in this same region, the
BLS linewidth increases by more than a factor of ten. This two opposit
behaviors indicate that there are at least two different attenuation
mechanisms: {\it i)} One of dynamic origin dominant in the low $Q$
(low frequency) region, and {\it ii}) A second one, dominant at high $Q$,
whose temperature independence suggests that its origin is due to the
structural disorder of the glass. 
The glycerol results are confirmed by a similar sound attenuation
behavior found in vitreous silica, as obtained by the analysis of
existing BLS, IXS and MD data. The observation of two distinct
attenuation mechanisms, each one dominant in a different $Q$ region,
implies the existence of a cross-over frequency, which lies in the
100 $GHz$ range for both of the studied glasses. It also suggests
that the frequency dependence of the dynamic contribution to the sound
attenuation agrees with the one recently predicted by Fabian and Allen
\cite{FA}.

The experiment has been carried out at the new very high energy resolution 
IXS beamline ID28, at the European Synchrotron Radiation Facility. The
incident x-ray beam is obtained by a back-scattering monochromator operating
at the Si(11~11~11) reflection \cite{noiV}. The scattered photons are
collected by a spherical silicon crystal analyzer, also operating at the
Si(11~11~11) reflection \cite{noiM}. The monochromatic beam has an energy
of $\approx$ 21,748~eV and an intensity of 2$\cdot$10$^8$~photons/s. 
The total energy resolution $-$ obtained from the measurement of $S(Q_M,E)$
in a Plexiglas sample which is dominated by elastic scattering $-$ is 1.5~meV
full-width-half-maximum (fwhm). The momentum transfer,
$Q=2k_{\circ}sin(\theta_s/2)$, with $k_{\circ}$ the wavevector of the
incident photon and $\theta_s$ the scattering angle, is selected between
2 and 4~nm$^{-1}$ by rotating a 7~m long analyser arm in the horizontal
scattering plane. The total $Q$ resolution has been set to 0.2~nm$^{-1}$.
Energy scans are done by varying the relative temperature between the
monochromator and analyzer crystals. Each scan took about 180~min,
and each $(Q,T)$-point spectrum has been obtained from the average
of 2 to 8 scans depending on the sample temperature. 
The data have been normalized to the intensity of the incident beam.
The sample cell is made out of a {\it pyrex}-glass tube (4~(10)~mm
inner (outer) diameter and 20~mm length), capped with two diamond
single crystals discs, 1~mm thick, to minimize undesired scattering
signals. The cell has been loaded with high purity glycerol in an
argon glove box. In the $Q-E$ region of interest, empty cell measurements
gave the flat electronic detector background of 0.6~counts/min.
The cell length was chosen to be comparable to 
the x-ray photoabsorption length, and multiple scattering 
was negligible.

The spectra have been collected at $T$=16, 45, 75, 114, 145 and 167 K,
and, as examples, those at $T=16$ and 167 K are reported in Fig.~1 for 
different $Q$-values. The full lines are the fits to the 
data, obtained using a model function made by the convolution of 
the experimentally determined resolution function with a delta 
function for the elastic peak and a Damped Harmonic Oscillator (DHO) 
model for the inelastic peaks \cite{sci}. This model for the $S(Q,E)$
results from the assumption that the memory function, $m_Q(t)$
\cite{boonyip}, entering in the Langevin equation for the considered
$Q$-component of the density fluctuation, has a time dependence as: 
$m_Q(t)=2 \Gamma_Q(T) \delta(t) + \Delta_Q^2(T)$ \cite{giulio}. 
The presence of the structural $\alpha$-relaxation, observed in the
liquid state and frozen in the glass, and of other relaxation processes 
with characteristic times slower than $\approx 1$ ps, is reflected in the
parameter $\Delta_Q(T)$, whose value determines the change of the sound 
velocity, $c$, between the fully relaxed ($c_o$) and unrelaxed ($c_\infty$) 
limiting values: $\Delta^2_Q(T)=Q^2(c^2_\infty-c^2_o)$. The parameter
$\Gamma_Q(T)$ determines the width of the side peaks, i.~e. the sound
wave attenuation coefficient, $\alpha=2\pi\Gamma_Q/h c$.
However, fits made with the DHO model or with different fitting function
gave values for the FWHM of the inelastic peaks consistent among each
other within their statistical uncertainties, indicating, therefore, the
insensibility of the results to the specific model for the inelastic
peaks. As it is evident
already from the raw data, $\Gamma_Q(T)$ has a marked $Q$-dependence 
while its $T$-dependence, if any, is much smaller.
This is better seen by the dotted lines in Fig.~1, which represent the
unconvoluted inelastic part of $S(Q,E)$.

The values of $\Gamma_Q(T)$ resulting from the fit of the 
IXS data of Fig.~1 are reported as a function of $Q$ in Fig.~2.
In the same figure are also shown the $\Gamma_Q(T)$ obtained 
from IXS measurements at 175 K \cite{sci}, and those obtained 
from literature BLS spectra measured at $Q\approx$0.03 nm$^{-1}$ 
and temperatures similar to the IXS' ones \cite{BLSgly,BLSgly2}. 
This figure demonstrates that, within the error bars, $\Gamma_Q(T)$ is 
{\it T-independent} in the $Q$ region covered by IXS. On the 
contrary, $\Gamma_Q(T)$ shows a marked $T$-dependence at the 
$Q$ value of the BLS measurements. These $T$-dependencies are 
emphasized in Fig.~3, where $\Gamma_Q(T)/Q^2$ is plotted as a function
of $T$ for $Q$ in the BLS region ($Q\approx 0.035$ nm$^{-1}$)
\cite{BLSgly,BLSgly2} and for $Q$ in the IXS region. Here the crossed
symbols refer to IXS measurements at fixed $Q$ ($Q=2$ nm$^{-1}$) 
and the full symbols to the average of $\Gamma_Q(T)/Q^2$ over the 
$Q=2-4$ nm$^{-1}$ region.
This figure confirms that, at the high $Q$ values, $\Gamma_Q(T)$
is substantially constant even at very low $T$, whereas, at the 
low $Q$ values, it increases with temperature up to $\approx 100$ K,
where it seems to reach a plateau. The further increase above $T_g$
is due to the $\alpha$-relaxation; it is only seen in the low $Q$ data 
as it would affect the high $Q$ data at higher $T$.

The specific dynamic mechanisms (anharmonicity, relaxation processes, 
floppy modes, two level systems...) at the origin of the acoustic
attenuation observed at the BLS' $Q$ values, as well as their
temperature dependence, have been widely investigated in the past
\cite{BLShun}.  In contrast to what it is found in the BLS' $Q$ region,
the behavior of $\Gamma_Q(T)$ in the high $Q$ region, as reported in 
Figs.~2 and 3, shows that here the sound attenuation is not determined
by temperature activated dynamic processes. Consequently, in this 
$Q$ range, at variance with the crystalline state where the absence 
of dynamic processes would imply no sound attenuation, in the glass
the observed non vanishing value of $\Gamma_Q$ must have a "structural" 
origin, i.~e. it must be due to the topological disorder of the glass 
structure. 

The picture coming from the reported data suggests that one can 
express $\Gamma_Q(T)=\Gamma^{_{(D)}}_Q(T)+ \Gamma^{_{(S)}}_Q$,
where $\Gamma^{_{(D)}}_Q(T)$ is a temperature dependent dynamic 
part and $\Gamma^{_{(S)}}_Q$ is due to topological disorder. 
The $Q$-dependence of $\Gamma^{_{(D)}}_Q(T)$ must be such to be
the dominant term of $\Gamma_Q(T)$ at small $Q$, while it must 
be negligible at large $Q$. This behavior is consistent with a 
recent calculation of the dynamic (anharmonicity) contribution to 
the sound attenuation by Fabian and Allen \cite{FA}, who predict a 
$\Gamma_Q(T) \propto Q^2$ up to a $Q_c$ value, above which 
$\Gamma_Q(T) = const$. In the case of amorphous silicon, 
$Q_c$ has been calculated to be in the 0.1 nm$^{-1}$ range \cite{FA}. 
Although there is no such calculation for glycerol, the present 
results indicate that also in this glass the crossover takes place 
at $Q$ values between the BLS' and IXS' ones, i.~e. in the 
$\nu_c\approx$100 GHz frequency range. It is worth to note that a 
relaxation process with characteristic time $\tau=1/2\pi\nu_c\approx 2$ 
ps, and responsible for a linewidth of the order $\Gamma_Q/Q^2\approx 0.2$ 
meV/nm$^{-2}$, should also give a dispersion of the sound velocity,
$\delta c$, given by $\delta c/c=\pi\Gamma_Q/ Q^2 \tau h c^2\approx$
1\%, a value too small to be detectable with the accuracy achievable 
at present. Therefore, in the present case, the change of sound velocity 
cannot be used to estimate the crossover value $Q_c$.

The $Q$-dependence of $\Gamma^{_{(S)}}_Q$, as already observed before in 
many other glasses and glass forming systems \cite{sci}, is well
represented by a $Q^2$ law, $\Gamma^{_{(S)}}_Q=D \; Q^2$, in the $Q$
region covered by IXS. This $Q^2$ law, shown as a full line in Fig.~2,
however, cannot be extrapolated (thin full line) to low $Q$ values because
it would predict width values in excess to the measured $\Gamma_Q(T)$.
This observation excludes that $\Gamma^{_{(S)}}_Q \propto Q^2$ in the
whole 0.01-10 nm$^{-1}$ $Q$ range. Under the hypothesis that
$\Gamma^{_{(S)}}_Q=Q^\gamma$, and assuming that at the lowest measured
temperature $\Gamma^{_{(S)}}_Q \equiv \Gamma_Q$, one finds consistency
within the error bars of both the IXS data and the BLS low temperature
point, as shown by the dashed line in Fig.~2 obtained with $\gamma=2.5$.
This estimate of $\gamma$ is a low limiting value, because the BLS low $T$
width could still be partially affected by a dynamic contribution. 
It is not clear, however, whether the hypothesized power law indeed provides
a good representation of $\Gamma^{_{(S)}}_Q$, some hints on this issue can
be gathered by the study of other glasses.

The previous picture for the sound attenuation in the glycerol glass is
further substantiated by the existing IXS \cite{IXSsio1,IXSsio2,IXSsio3}, 
BLS \cite{BLSsio1,BLSsio2,BLSsio3,BLSsio4,BLSsioA}, 
MD \cite{siomd,MDsioE} and Picosecond Optical Technique (POT) \cite{POT}, 
data in another prototypical glass: vitreous silica (v-$SiO_2$). We report 
in Figs.~4 and 5 the $\Gamma_Q(T)$ values for v-$SiO_2$ in a format
equivalent to that of Figs.~2 and 3. In Fig.~4, the $\Gamma_Q(T)$ values
in the 1 to 5 nm$^{-1}$ region, as obtained by IXS, do not show any
relevant $T$-dependence in the 300-1450 K range. As a consequence of
contrast problems due to the limited energy resolution of the IXS
spectrometer, at temperatures below 300 K, it has not been possible
to discriminate the inelastic signal from the tails of the elastic one.
The missing low temperature IXS data are supplied by an extended MD
simulation performed in the harmonic approximation ($T=0$ K), which
provides a $S(Q,E)$ lineshape in excellent agreement with the IXS data,
and confirms the absence of any relevant $T$-dependence of the sound
attenuation in the whole $T$=0-1450 K and $Q$=1-5 nm$^{-1}$ ranges.
On the contrary, as in glycerol and as emphasized in Fig.~5, the BLS
data of v-$SiO_2$ data show a large $T$-dependence. As in other glasses,
also the high $Q$ v-$SiO_2$ data of $\Gamma_Q$ have a $Q$-dependence
well represented by a $Q^2$ law (full line in Fig.~4). In v-$SiO_2$
are also available room $T$ data in a $Q$ region intermediate to BLS
and IXS ($Q$=0.03-0.4 nm$^{-1}$), as obtained by POT. These data,
however, show a relevant inconsistency with BLS data measured in
the same $Q$ and $T$ ranges.

The similar behavior between v-$SiO_2$ and glycerol allows to
formulate also for v-$SiO_2$ the same hypotheses on the $Q$ and
$T$-dependencies of $\Gamma^{_{(D)}}_Q(T)$ and $\Gamma^{_{(S)}}_Q$.
It is worth to note, however, that the ability of the power law
for $\Gamma^{_{(S)}}_Q$ (which should have $\gamma=2.6$) to pass
through all the IXS data and the low $T$ BLS point is substantially
worse than in glycerol. Therefore the v-$SiO_2$ results clearly
indicates that the power law hypothesis is wrong and that
$\Gamma^{_{(S)}}_Q$ has a more complex behavior, namely it
is $\propto Q^2$ in the high $Q$ regime ($Q>1$ nm$^{-1}$)
and it has a steeper behavior at low $Q$ values.

In conclusion, we have shown that in glycerol and silica glasses the
dominant sound attenuation mechanism has a different origin in the $Q$
region spanned by the IXS or BLS techniques. The temperature independence
of the attenuation at large $Q$, corresponding to frequencies in the 
{\it THz} region, implies that its origin is structural and it is due to 
the disorder. On the contrary, the well known strong $T$-dependence found 
in the low $Q$ region, at {\it GHz} frequencies and below, implies a dynamic 
origin of the sound attenuation \cite{BLShun}. These findings imply a
cross-over region between the two regimes, which should lie in the 100
GHz frequency range. The overall $T$ and $Q$-dependencies of the attenuation
considered here is consistent with a dynamic ($\Gamma^{^{(D)}}_Q$) part
that closely follows the one proposed by Fabian and Allen \cite{FA}, and
a structural part ($\Gamma^{_{(S)}}_Q$) that has a $Q^2$ behavior at high
$Q$ ($Q>1$ nm$^{-1}$) and a steeper $Q$ dependence at lower $Q$. The
hypothesis on a cross-over between the two different attenuation
mechanisms considered here call for further studies,
where the $Q$ and $T$-dependencies of the sound waves is thoroughly
investigated using the BLS and IXS technique in a wider number of
glass materials.

{\footnotesize{
\begin{center}
{\bf FIGURE CAPTIONS}
\end{center}
\begin{description}
\item  {FIG. 1 -
Inelastic x-ray scattering spectrum of glycerol at T=16 and 167 K 
and at the indicated $Q$-values. The full lines are the best fit to the 
data as discussed in the text. The dotted lines represent the uncovoluted 
inelastic contributions to the fit. 
}
\item  {FIG. 2 -
The linewidth parameters $\Gamma_Q$ for glycerol are reported as a function 
of $Q$ at the indicated temperatures in the IXS' (full symbols) and BLS' 
(open symbols) $Q$ regions. The inset shows an enlargement of the IXS' $Q$
region. The full line represents the $Q^2$ behavior, which is the best 
fit to $\Gamma_Q$ at high $Q$. Also shown in the figure are extrapolation
of the $Q^2$ law in the low $Q$ range (thin full line) and the $Q^{2.5}$ 
dependence (dashed line) indicated by the low $T$ BLS data. The inset shows
an enlargement of the high $Q$ region.
}
\item  {FIG. 3 -
Temperature dependence of $\Gamma_Q/Q^2$ in glycerol at $Q\approx0.03$
nm$^{-1}$ (open symbols), at $Q=2$ nm$^{-1}$ (crossed symbols) and
averaged over the $Q=2-4$ nm$^{-1}$ region (full symbols). The vertical
dashed line indicates the glass transition temperature, $T_g$=187 K. 
}
\item  {FIG. 4 -
Same as in Fig.~2 but for vitreous silica. Data from MD simulation (crossed
symbols) and from POT (stars) are also reported. Here the low $T$ BLS data
indicate (if any) a power law with exponent $\gamma=2.6$. The inset shows an
enlargement of the high $Q$ region.
}
\item  {FIG. 5 -
Same as in Fig.~2 but for vitreous silica. The inset shows an enlargement
of the low temperature region. The open circle refer to BLS data at
$Q\approx 0.035$ ($\Box$ and $\circ$) and $Q\approx 0.025$ ($\Diamond$)
nm$^{-1}$, the full squares to IXS data at $Q=1.6$ nm$^{-1}$, and the
crossed circle to MD simulation in the harmonic ($T$= 0 K) limit. The
inset shows an enlargement of the low $T$ region.
}
\end{description}
}}
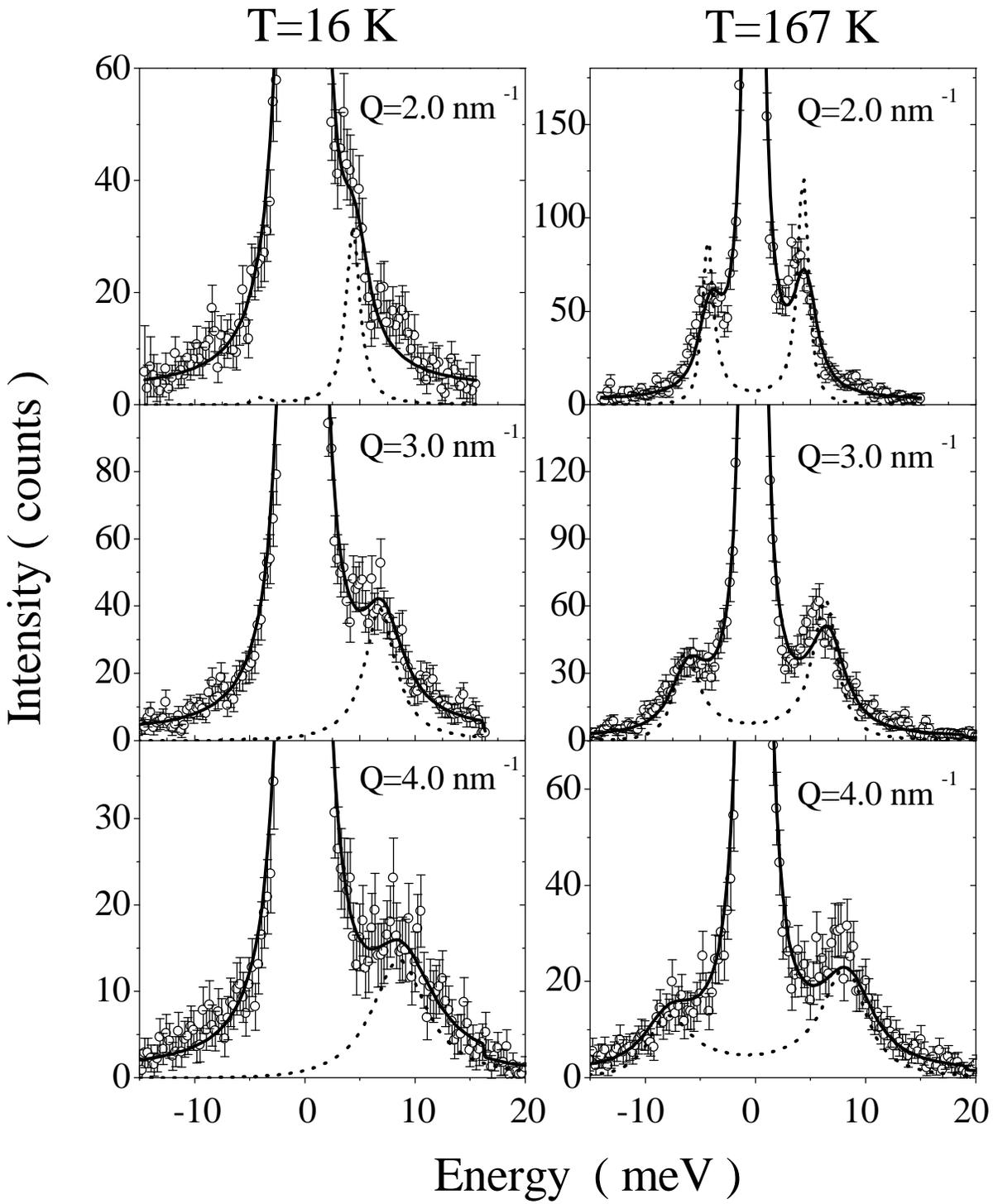
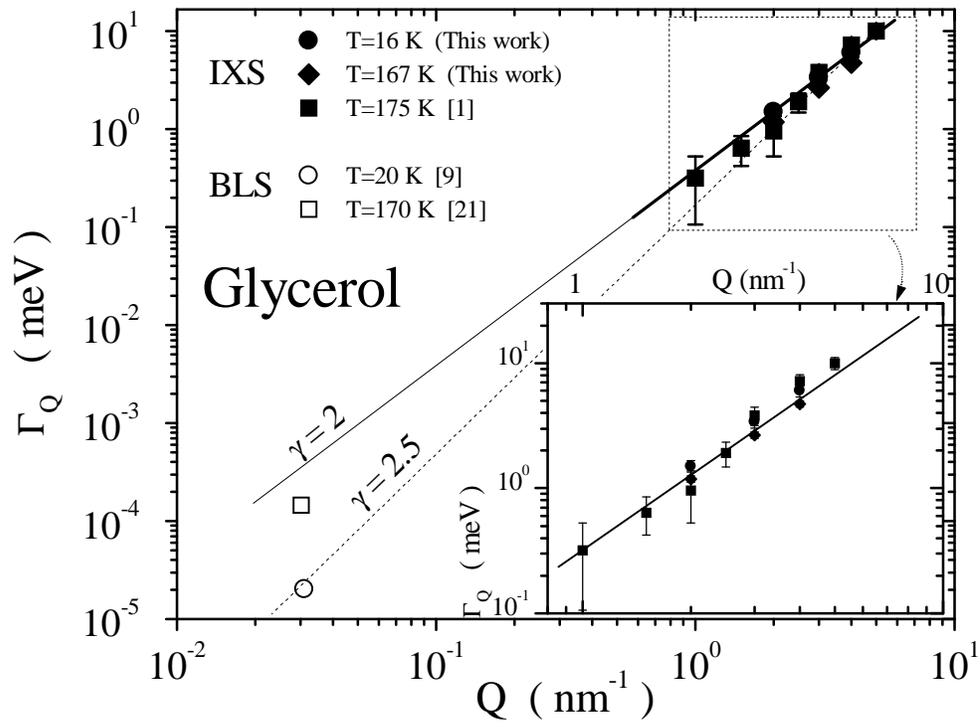
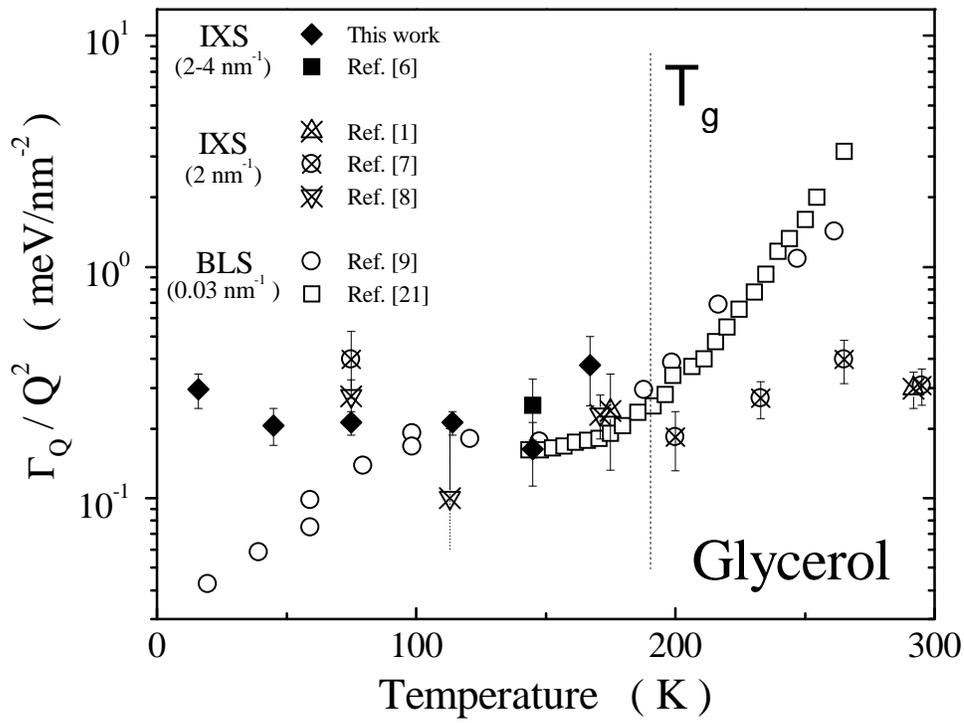
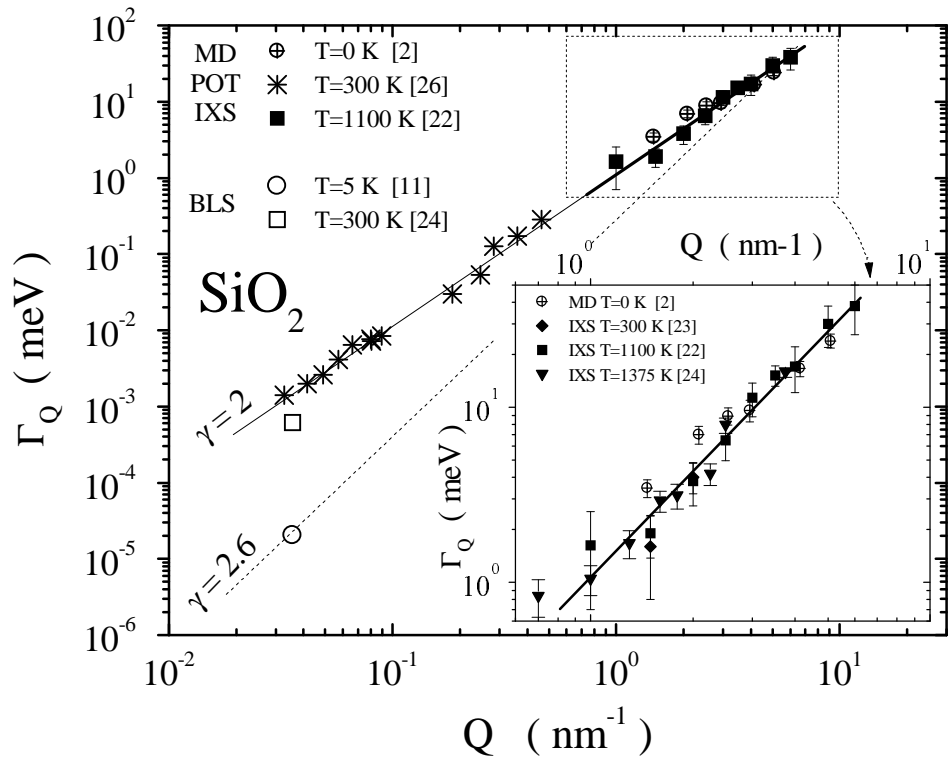
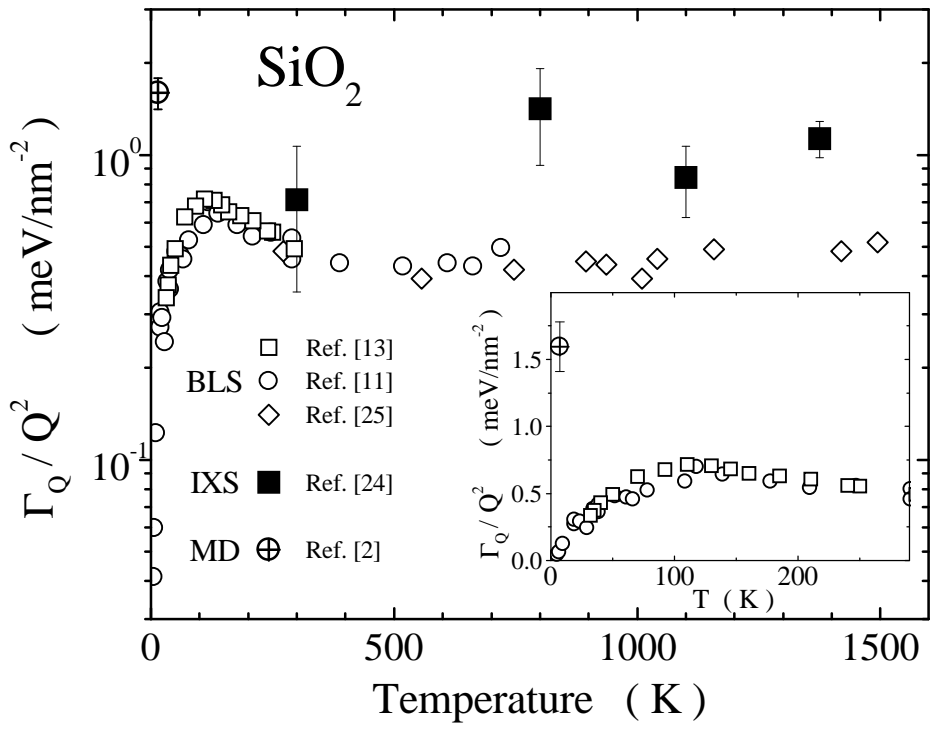
\end{document}